# A Density Functional Study of Atomic Hydrogen and Oxygen Chemisorption on the Relaxed (0001) Surface of Double Hexagonal Close Packed Americium

Pratik P. Dholabhai, Raymond Atta-Fynn, and Asok K. Ray*

Physics Department, University of Texas at Arlington, Arlington, Texas 76019

*akr@uta.edu.

**Abstract**

*Ab initio* total energy calculations within the framework of density functional theory have been performed for atomic hydrogen and oxygen chemisorption on the (0001) surface of double hexagonal packed americium using a full-potential all-electron linearized augmented plane wave plus local orbitals method. Chemisorption energies were optimized with respect to the distance of the adatom from the relaxed surface for three adsorption sites, namely top, bridge, and hollow hcp sites, the adlayer structure corresponding to coverage of a 0.25 monolayer in all cases. Chemisorption energies were computed at the scalar-relativistic level (no spin-orbit coupling NSOC) and at the fully relativistic level (with spin-orbit coupling SOC). The two-fold bridge adsorption site was found to be the most stable site for O at both the NSOC and SOC theoretical levels with chemisorption energies of 8.204 eV and 8.368 eV respectively, while the three-fold hollow hcp adsorption site was found to be the most stable site for H with chemisorption energies of 3.136 eV at the NSOC level and 3.217 eV at the SOC level. The respective distances of the H and O adatoms from the surface were found to be 1.196 Å and 1.164 Å. Overall our calculations indicate that chemisorption energies in cases with SOC are slightly more stable than the cases with NSOC in the 0.049-0.238 eV range. The work functions and net magnetic moments respectively increased and decreased in all cases compared with the corresponding quantities of bare dhcp Am (0001) surface. The partial charges inside the muffin-tins, difference charge density distributions, and the local density of states have been used to analyze the Am-adatom bond interactions in detail. The implications of chemisorption on Am 5*f* electron localization-delocalization are also discussed.

**1 Introduction**

Surface chemistry and physics have been and continues to be very active fields of research because of the obvious scientific and technological implications and consequent importance of such research. One of the *many* motivations for this burgeoning effort has been the desire to understand surface corrosion, metallurgy and catalytic activity in order to address environmental concerns. In particular, such efforts are important for a group of strongly correlated and heavy fermion systems like the actinides, for which experimental work is relatively difficult to perform due to material problems and toxicity [1-5]. Radioactive and highly electropositive, the actinides are characterized by the gradual filling of the 5*f* electron shell with the degree of localization increasing with the atomic number Z along the last series of the periodic table. The open shell of the 5*f* electrons determines the magnetic and solid state properties of the actinide elements and their compounds. However, these properties of the actinides, particularly the transuranium actinides, are *still not* clearly understood. This stems primarily from the inherent difficulty in understanding the behavior of the 5*f* electrons, whose spatial extent and tendency to interact with electrons on ligand sites gives rise to the chemically complex nature of the transuranium actinides. The actinides are also characterized by the increasing prominence of relativistic effects and their study can, in fact, give us an in depth understanding of the role of relativity throughout the periodic table.

Among the transuranium actinides, the unique electronic properties of the manmade Americium (Am) metal, which was first successfully synthesized and isolated at the wartime Metallurgical Laboratory [6], have received increased interests recently, from both scientific and technological points of view. Am occupies a pivotal position in the actinide series with regard to the behavior of *5f* electrons [7]. Atomic volumes of the actinides as a function of atomic number have experimentally displayed a sharp increase between Pu and Am [8] . In contrast to this sharp increase, the atomic volumes of the actinides before Pu continuously decreases as a function of increasing atomic number from Ac until Np, a behavior analogous to *d* transition metals. These behaviors reveal that the properties of the *5f* electrons change dramatically starting from somewhere between Pu and Am. It has been suggested [9, 10] that the *5f* electrons of the actinides before Am participate in bonding while the *5f* electrons of the actinides after Pu become localized and non bonding. As a result, several experimental and theoretical works have been done in recent years to gain insight into the structural and electronic properties of Am [11-25]. Central to the questions concerning Am are the phase transitions with increasing pressure, localization/delocalization behavior of the *5f* electrons, and *possible* magnetism of Am. We have discussed these issues and the relevant literature, in detail, in our previous work on the quantum size effects in fcc and dhcp Am [26]. In particular, the anti-ferromagnetic state with spin-orbit coupling was found to be the ground state of dhcp Am with the *5f* electrons primarily localized and the surface energy and work function of the of the dhcp Am(0001) surface were predicted to be 0.84 J/m$^2$ and 2.90eV. As a continuation of our systematic density functional studies of adsorption processes of environmental gases on actinide surfaces [27], we report in this work, detailed *ab initio* electronic and geometric structure studies of atomic hydrogen and oxygen adsorbed on the (0001) surface of dhcp Am. To the best of our knowledge, *no* such study exists in the literature, though, as mentioned in our previous works, an effective way to probe the actinides' *5f* electron properties and their roles in chemical bonding is a study of their bare surface properties and atomic and molecular adsorptions on them.

## 2 Computational method

All calculations have been performed within the generalized gradient approximation to density functional theory (GGA-DFT) with the Perdew-Burke-Ernzerhof (PBE) exchange-correlation functional [28, 29]. The Kohn-Sham equations were solved using the full-potential linear augmented plane wave plus local basis (FP-LAPW+lo) method as implemented in the WIEN2k code [30]. This method makes no shape approximation to the potential or the electron density. Within the FP-LAPW+lo method, the unit cell is divided into non-overlapping muffin-tin spheres and an interstitial region. Inside the muffin-tin sphere of radius $R_{MT}$, the wave functions are expanded using radial functions (solutions to the radial Schrödinger equation) times spherical harmonics with angular momentum up to $l_{max}^{wf}$=10. Non-spherical contributions to the electron density and potential inside the muffin tin spheres were considered up to $l_{max}^{pot}$=6. APW+lo basis were used to describe $s, p, d,$ and $f$ ($l$=0, 1, 2, 3) states and LAPW basis were used for all higher angular momentum states in the expansion of the wave function. Additional local orbitals (LO) were added to the 2$s$ semi-core states of O and the 6$s$, 6$p$ semi-core states of Am to improve their description. The radii of the muffin-tin spheres were $R_{MT}$ (H) = $R_{MT}$ (O)=1.2 Bohr and $R_{MT}$(Am) = 2.2 Bohr. The truncation of the modulus of the reciprocal lattice vector used for the expansion of the wave function in the interstitial region $K_{MAX}$, was set by $R_{MT}K_{MAX}$ = 8.5 for the clean slab and $R_{MT}K_{MAX}$ = 4.64 for the slab-with-adatom, where $R_{MT}$ denotes the smallest muffin tin radius, that is, $R_{MT}$ = 2.2 a.u. for the bare slab and $R_{MT}$ = 1.2 a.u. for the slab-with-adatom.

In the WIEN2k code, core states are treated at the fully relativistic level. Semi-core and valence states are treated at either the scalar relativistic level, i.e., no spin-orbit coupling (NSOC) or at the fully relativistic level, i.e., spin-orbit coupling (SOC) included. Spin-orbit interactions for semi-core and valence states are incorporated via a second variational procedure using the scalar relativistic eigenstates as basis [31], where all eigenstates with energies below the cutoff energy of 4.5 Ry were included, with the so-called $p_{1/2}$ extension [32], which accounts for the finite character of the wave function at the nucleus for the $p_{1/2}$ state. We considered both the NSOC and SOC levels of theory to investigate spin-orbit coupling effects on chemisorption energies.

The dhcp-Am (0001) surface was modeled by a supercell consisting of a periodic 6-layer slab with a (2×2) surface unit cell and vacuum of 30 Bohr thickness. In accordance with our previous findings [26], we have used an AFM configuration for the slab which consists of alternating ferromagnetic layers of up-spin and down-spin atoms along the $c$-axis. The spin quantization axis for the magnetic SOC calculations was along the [001] direction. The relaxation of the surface was carried out in two steps: first, bulk dhcp Am was optimized followed by surface optimization. The atomic volume of bulk dhcp Am was expressed in terms a single lattice constant by constraining the ratio $c/a$ to match experimental value. More precisely, the ratio $c/a$ was set to 3.2 (experimental ratio) and the volume V was expressed in terms of only $a$. Then the total energy $E$ (for an AFM configuration) was computed for several variations of $V$. The energy versus volume E-V fit via Murnaghan's equation of state [33] generated an equilibrium volume Vo = 208.6 (a.u.)$^3$ and B = 25.4 GPa. The equilibrium volume Vo corresponded to $a$=6.702 a.u. The experimental values are 198.4 (a.u.)$^3$ or 197.4 (a.u.)$^3$

and 29.7 GPa [34, 35]. Integrations in the Brillouin zone (BZ) have been performed using the special k-points sampling method with the temperature broadening of the Fermi surface by the Fermi distribution, where a broadening parameter $K_BT = 0.005$ Ry has been used. The temperature broadening scheme avoids the instability from level crossings in the vicinity of the Fermi surface in metallic systems and also reduces the number of k-points necessary to calculate the total energy of metallic systems [36]. For the present work, a 6×6×1 k-mesh density (18 k-points in the irreducible part of the BZ) was deemed to be sufficient. Self-consistency is achieved when the total energy variation from iteration to iteration is 0.01 mRy or better. Using the optimized lattice constants, that is, $a$=6.702 a.u. and c = 3.2$a$, a 2x2 hexagonal surface unit cell (2 atoms along each lateral 2D direction yielding 4 Am atoms per surface unit cell) for (0001) orientation is constructed. Then the surface unit cell is used to build the slab with 6 atomic layers (with the proper layer stacking *ABABAB*....., taken into account) and 30 a.u. vacuum. Furthermore, the slab was built to have inversion symmetry for computational efficiency. The interlayer spacing between the surface unit cells in the slab above corresponded to the bulk spacing $d_0$= *c/4*. Next, the central layers are fixed at the bulk positions but the 2 outermost layers (this is the same from both sides of the central slabs because of inversion symmetry) are allowed to relax in order to lower the total energy. The relaxation was performed by minimizing the total energy by varying $d_{12}$, the separation between the central and subsurface layers and $d_{23}$. the separation distance between the subsurface and surface layers (variations of -4%, -2%, 0%, 2%, 4% measured in terms of the bulk interlayer spacing $d_0$for $d_{12}$ and $d_{23}$ resulting in a 5x5 grid for the energy computation). The relaxations obtained were $\Delta d_{12}/d_0 = 0\%$ and $\Delta d_{23}/d_0 = 2\%$, where $d_0$ is the bulk interlayer separation, with the reduction in the total energy of the slab being 2.19 mRy. The small relaxations and reduction in the total energy indicate the fair stability of the surface. To study adsorption on the relaxed Am surface, the adatom, corresponding to a surface coverage of $\Theta = 1/4$ ML, was allowed to approach the surface from both sides to preserve inversion symmetry. Three high symmetry adsorption sites were considered (see Fig. 1): (i) one-fold top site (adatom is directly on top of a Am atom) (ii) two-fold bridge site (adatom is placed in the middle of two nearest neighbor Am atoms); and (iii) three-fold hollow hcp site (adatom "sees" a Am atom located on the layer directly below the surface layer). The chemisorption energy $E_C$ is optimized with respect to the height R of the adatom above the bare relaxed surface. No *further* surface relaxations and/or reconstructions were taken account for both physical (any further relaxations is expected to be quite small) and computational reasons We believe though that further relaxations and/or reconstructions during adsorption, *if any,* will *not* significantly alter the results reported here both qualitatively *and* quantitatively. The chemisorption energy $E_C$ is given by:

$$E_C(R) = 1/2[E(Am) + 2E(X) - E(Am+X)],$$

where E(Am) is the total energy of the bare Am slab, E(X) is the total energy of the isolated adatom, and E(Am+X) is the total energy are the slab-with-adatom. A positive value of $E_C$ implies chemisorption and a negative value implies otherwise. To calculate the total energy of the adatom, the isolated atom was simulated in a large box of side 25 Bohr and at the Γ k-point, with all other computational parameters remaining the same. Also, our recent studies on adsorption on the δ-Pu surface indicated that spin-orbit coupling has negligible effect on the adsorption geometry but the binding was slightly stronger with the chemisorption energies increasing by 0.05 to 0.3 eV. Though we have not verified this explicitly for the dhcp Am (0001) surface, we expect the same trend to hold here. Hence in the current calculations, the geometry was optimized at the NSOC level and the final geometry was used for the SOC calculation to study the effects of spin-orbit coupling on the adsorption energies.

### 3 Results and discussions

Table 1 lists the adsorption energies and associated geometrical information of the H and O atoms adsorbed on the (0001) surface of dhcp-Am. The differences between the NSOC and SOC chemisorption energies at each adsorption site, given by $\Delta E_C = E_C(SOC) - E_C(NSOC)$, are also listed. For H adsorption, the trend in the chemisorption energies at the NSOC level of theory is the same as the SOC case. The most stable site is the hollow hcp site (3.136 eV for the NSOC case, 3.217 eV for SOC case), closely followed by the bridge adsorption site (2.965 eV for NSOC case, 3.014 eV for the SOC case), with the least favorable site being the top site (2.272 eV for the NSOC case, 2.377 eV in the SOC case). The vertical height R of the H atom above the surface layer clearly show that at the least stable top site, the adatom is furthest away from the surface (2.122 Å) followed by the next stable bridge site (1.429 Å), with the vertical height of the adatom from the surface layer being the lowest at the most stable hollow hcp site (1.196 Å). Hence, increasing stability at both the NSOC and SOC levels of theory implies decreasing vertical distance of the H adatom from the surface layer. Also increasing stability implies increasing adatom coordination number at both theoretical levels, that is, the H adatom prefers to bind at the maximally coordinated three-fold hollow hcp site. The Am-H bond lengths listed in Table I also indicate a relationship with the adatom coordination numbers, with the one-fold coordinated top site having the shortest bond and the three-fold hollow hcp site having the longest bond. All chemisorption energies indicate that binding is slightly stronger with the inclusion SOC compared the NSOC case. The SOC-NSOC chemisorption energy differences $\Delta E_C$ are listed in Table I; $\Delta E_C$ is maximum at the least stable top site (0.105 eV) closely followed by the next stable hollow hcp adsorption site (0.081 eV), with the intermediately stable bridge adsorption site having an SOC-NSOC $\Delta E_C = 0.049$ eV.

For O adsorption, the trend in the chemisorption energies for the NSOC case is also the same as that in the SOC case. The most stable site is the bridge site (8.204 eV for the NSOC case, 8.368 eV for SOC case). This is closely followed by the hollow hcp site (8.109 eV for NSOC case, 8.347 eV for the SOC case), with the least favorable site being the top site (6.388 eV for the NSOC case, 6.599 eV in the SOC case). As a comparison, for NSOC calculations for O adsorption on the δ – Pu (111) surface, the hollow hcp adsorption site was found to be the most stable site for O with chemisorption energy of 8.025 eV. For SOC calculations, the hollow fcc adsorption site was found to be the most stable site with a chemisorption energy of 8.2 eV respectively. The optimized distance of the O adatom from the Pu surface was found to be 1.25 Å. The vertical height R of the O atom above the Am surface layer clearly show that for the least stable top site, the adatom is furthest away from the surface (1.911 Å), followed by the most stable bridge site (1.164 Å), with smallest distance corresponding to the intermediately stable hollow hcp site (0.878 Å). Here, unlike the case for H, increasing stability at both the NSOC and SOC levels of theory does not necessarily imply decreasing vertical distance of the O adatom from the surface layer, since the distance at the most stable bridge site is greater than the distance at the next stable hollow hcp site. Furthermore, the most preferred bridge adsorption site does not have the maximum coordination. However, chemisorption energies in the SOC case are more stable than the NSOC case; $\Delta E_C$ is maximum at the hcp hollow site (0.238 eV) closely followed by the least stable top adsorption site (0.211 eV), with the most stable bridge adsorption site having an SOC-NSOC $\Delta E_C$ = 0.164 eV. Also, it is worth noting that the Am-H bond lengths are longer than Am-O bond lengths for each adsorption site as expected.

In table 2, the adsorbate-induced work function changes with respect to the clean metal surface, given by $\Delta\Phi = \Phi^{adatom/Am} - \Phi^{Am}$, are listed for the NSOC and SOC levels of theory for each adsorbate and each adsorption site. We observe that for the hydrogen adatom at each theoretical level high chemisorption energies generally correspond to low work function shifts. In fact, the changes in the work functions are largest at the least preferred top site and lowest at the most preferred hcp hollow site. This is not true for oxygen adsorption though, where the most preferred bridge site has a higher change in the work function compared to the hcp hollow site. In both cases however, the magnitude of the work function shifts is related to the adsorption site coordination; the lower coordinated top site shows the largest shift and the higher coordinated hollow hcp site shows the lowest shift. The adsorbate-induced work function shifts can be understood in terms of the surface dipoles arising due to the displacement of electron density from the substrate towards the adsorbates since the electronegativities of the adsorbates are larger than that of Am. The surface dipole moment μ (in Debye) and the work function shift $\Delta\Phi$ (in eV) are linearly related by the Helmholtz equation $\Delta\Phi = 12\Pi\Theta\mu/A$, where A is the area in Å² per (1×1) surface unit cell and Θ is the adsorbate coverage in monolayers. From the Helmholtz equation, we see that for each adsorbed adatom, μ is largest at the top site and smallest at the hcp hollow site.

In table 3, the magnitude and alignment of the site projected spin magnetic moments for each Am atom on each atomic layer, as well as the net spin magnetic moment per Am atom is reported for the clean metal surface and the chemisorbed systems. Here we report the moments for the SOC calculations. NSOC moments follow a similar qualitative trend and are not reported here. $\mu_1, \mu_2$, and $\mu_3$ are respectively moments for the surface, subsurface, and central layers. The spin moments for each of the four Am atoms on each layer are reported. $\mu_{int}$ and $\mu_{tot}$ are the interstitial spin moment and net moment per Am atom, respectively. First, it is clearly evident that the values of $\mu_2$ and $\mu_3$ (site spin moments in all for subsurface and central layers) in the chemisorbed systems is virtually the same as that of the clean slab and the major changes in the spin magnetic moments occurring mainly on the surface layer. As a result, all discussions regarding the spin magnetic moments will be confined to the surface layer. For each adsorption site, the spin moment of the closest neighbor surface layer Am atoms with which the adatom primarily interacts is indicated in bold fonts in the table 3. For the top sites, we see reductions of 0.14 $\mu_B$ and 0.62 $\mu_B$ in the spin moment of the Am atom for H and O chemisorptions respectively, while the moments of the remaining three Am atoms remain basically unaltered when compared to the clean surface. This naturally leads to a reduction in the net spin magnetic moment per Am atom. For the bridge sites, we see reductions of 0.08 $\mu_B$ and 0.27 $\mu_B$ in the spin moments of each of the two surface

Am atoms for H and O chemisorptions respectively, while very small or no change in the moments of the other two Am atoms occurs when compared to the clean slab. For the hollow hcp sites, reductions of 0.06 $\mu_B$ and 0.16 $\mu_B$ in the spin moments for each of the three Am atoms can be observed for H and O respectively, with little or no changes in the moment of the fourth Am atom. Also, the moments in the interstitial region $\mu_{int}$ decrease upon chemisorption. Hence the reduction in the net moment stems primarily from the reduction in the spin moments of the surface Am atoms interacting with the adatoms.

Due to the nature of the APW+lo basis, the electronic charges inside the muffin-tin spheres can be decomposed into contributions from the different angular momentum channels. We refer to these charges as partial charges. By comparing the partial charges $Q_B$ of the Am layers and adatoms before adsorption to the corresponding partial charges $Q_A$ after adsorption, we can get an idea of the nature of the interaction between the adsorbate and substrate. Thus we have reported $Q_A$ and $Q_B$ for each adatom and the Am atoms at each adsorption site in tables 4-9. In each table, we have also reported the differential partial charge of a given state $l$ corresponding to a given atom, which is given by $\Delta Q(l) = Q_A - Q_B$. $\Delta Q(l) > 0$ indicates charge gain inside the muffin tin sphere while $\Delta Q < 0$ indicates otherwise. First, it is worth noting that $\Delta Q(l)$ is quite small (±0.01) or completely vanishes for the Am atoms on the subsurface and central layers, with "significant" changes occurring on the surface layer. Hence the partial charges will be discussed only for the surface layer. In tables 4-6, $Q_A$, $Q_B$, and $\Delta Q(l)$ for H adsorbed at the top, bridge, and hcp sites respectively on the dhcp-Am(0001) surface are reported. For the one-fold top site (table 4), $\Delta Q(1s) = 0.16$ for H, $\Delta Q(6d) = 0.04$ and $\Delta Q(5f) = -0.09$ for the Am atom, implying significant Am(6d)-Am(5f)-H(1s) hybridizations. For the two-fold bridge site (table 5), $\Delta Q(1s) = 0.18$ for H, $\Delta Q(6d) = 0.01$ and $\Delta Q(5f) = -0.03$ for each of the two Am atoms, suggesting the participation of some of the Am $5f$ electrons in chemical bonding with H. For the three-fold hollow hcp site (table 6), $\Delta Q(1s) = 0.18$ for H, $\Delta Q(6d) = 0.01$ and $\Delta Q(5f) = -0.02$ for each of the three Am atoms, which again suggest some contribution of the Am $5f$ electrons to Am-H bonding. In tables 7-9, $Q_A$, $Q_B$, and $\Delta Q(l)$ for O adsorbed at the top, bridge, and hcp sites respectively on the dhcp-Am(0001) surface are reported. For the top site (table 7), $\Delta Q(2p) = 0.27$ for O, $\Delta Q(6d) = 0.21$ and $\Delta Q(5f) = -0.24$ for the Am atom, which like the case for H, implies significant Am($6d$)-Am($5f$)-O($2p$) interactions. For the bridge site (table 8), $\Delta Q(2p) = 0.25$ for O, $\Delta Q(6d) = 0.06$ and $\Delta Q(5f) = -0.11$ per each of the two Am atoms, suggesting the participation of some the Am $5f$ electrons in Am-O bonding. For the hollow hcp site (table 9), $\Delta Q(2p) = 0.27$ for O, $\Delta Q(6d) = 0.02$ and $\Delta Q(5f) = -0.06$ per each of the three Am atoms, which again suggest some contribution of the Am $5f$ electrons to Am-O chemical bonding. Overall, the partial charge analyses tend to suggest that some of the Am $5f$ electrons participate in chemical bonding. We wish to stress that the partial charges are confined inside the muffin tin spheres and do not give any information of the interactions between the atoms in the interstitial region. Information which includes the electronic charges in interstitial region can be obtained from the difference charge density distributions.

To investigate the nature of the bonds that have been formed between the adatoms and the Am atoms on the surface, we computed the difference charge density distribution. We define the difference charge density $\Delta n(r)$ as follows:

$$\Delta n(r) = n(X+Am) - n(Am) - n(X),$$

where n(X+Am) is the total electron charge density of the Am slab-with-X adatom, n(Am) is the total charge density of the bare Am slab, and n(X) is the total charge density of the adatom. In computing n(X) and n(Am), the adatom X and Am atoms are kept fixed at exactly the same positions as they were in the chemisorbed systems. All charge densities reported here were computed in the plane passing through the adatom and two surface Am atoms using the Xcrysden utility [37]. For the 1-fold coordinated top site, the plane passes through the adatom, the Am atom interacting with the adatom, and a neighboring Am atom. For the 2-fold coordinated bridge site, the plane passes through the adatom and the two Am atoms interacting with the adatom. For the 3-fold coordinated hollow hcp site, the plane passes through the adatom and the two of three Am atoms interacting with the adatom. In fig. 2, the difference charge densities distribution for H and O adsorptions are shown for each site. For the top site, we clearly see charge accumulation around each adatom and significant charge loss around the Am atom bonded to the adatoms, implying that the bond has a strong ionic character, which is expected as the adatoms are more electronegative than Am. Also the charge loss around Am is larger for O chemisorption since O is more electronegative than H. For the bridge and hollow hcp sites, the Am-adatom bonds are again largely ionic in character as significant charge accumulation around the adatoms can be observed. The difference charge density plots are fairly consistent with the differential partial charges reported in tables 4-9.

The local density of states, which is obtained by decomposing the total density of the single particle Kohn-Sham eigenstates into contributions from each angular momentum channel $l$ of the constituent atoms inside the muffin tin sphere have also been examined. Here, we have reported the LDOS for only the SOC computation as the DOS for NSOC calculations yields a similar qualitative description. In fig. 3, the Gaussian-broadened (with a width of 0.045 eV) $f$ and $d$ LDOS curves for each of the layers of the bare dhcp Am (0001) metal slab are shown. Clearly, we see well-defined peaks in the $5f$ electron LDOS in the vicinity of the Fermi level, which have also been observed for bulk dhcp-Am(0001), and is a clear signature of $5f$ electron localization [26], Also, the $5f_{5/2}$ electron localization is more pronounced for the surface and subsurface layers than the central layer. However, the $5f_{5/2}$ peak centered on a binding energy of 1 eV below the Fermi level instead of the 2.8 eV observed in X-ray and ultraviolet photoemission spectra experiments [12, 18].

In FIG. 4, we show the LDOS plots for the H adatom and the surface Am atoms before and after chemisorption. As there are four nonequivalent sites on the surface, we depict the LDOS for only the Am atom(s) directly bonded to the adatom in order to assess the changes in DOS upon chemisorption. At the top site, we note some modification in the Am $5f$ DOS just below the Fermi level in comparison to the $5f$ DOS before adsorption which implies that the some $5f$ electrons participate in chemical bonding. We also observe significant Am ($6d$)-H ($1s$) hybridizations with a small admixture of Am ($6f$) states, implying that the Am contribution to bonding is dominated by the $6d$ electrons. The LDOS distributions for the bridge and hcp hollow sites show a slight reduction in the $5f$ DOS below the Fermi level, with the H $1s$ bonding state pushed to lower binding energies, which naturally suggests stronger binding as observed in the chemisorption energies. Except for the slight reduction in the $5f$ DOS below the Fermi level, it is fair to say that the localization of the 5*f* bands are primarily retained after chemisorption.

In FIG. 5, the LDOS plots for O chemisorptions are shown. For the top site, we see a significant character of the 5*f* and 6*d* bands of Am in the O 2*p* bands. This indicates significant Am (6*d*)-Am (5*f*)-O(2*p*) hybridizations in the -4 eV to -2 eV energy range, which is in fair agreement with the partial charge analysis. The hybridizations lead to modifications in the 5*f* DOS below the Fermi level at the top sites. For the bridge and hcp sites, hybridizations is dominated by Am(6*d*)-O(2*p*) and only slight modifications in the 5*f* bands below the Fermi level is observed. The overlap of O 2*p* bands with the Am 5*f* bands decreases and O 2*p* bands are pushed to lower energies which is reflected in the strong binding energies. Just like H adsorption, the sharp and peaky nature of the 5*f* bands are retained in general after chemisorption as no significant broadening of the bands is observed.

## 4. Conclusions

In summary, we have used the generalized gradient approximation to density functional theory with the full potential LAPW+lo method to study chemisorption of H and O atoms on the (0001) surface of dhcp Am at two theoretical levels; one with no spin-orbit coupling (NSOC) and the other with spin-orbit coupling (SOC). For H adsorption, the hollow hcp site was the most preferred site, while the bridge adsorption was the most preferred site in O adsorption. The inclusion of spin-orbit coupling lowers the chemisorption energies by 0.049-0.238 eV. Work functions increased in all cases compared to the clean Am surface, with the largest shift corresponding to the least coordinated top site and lowest shifts corresponding to the maximally coordinated hollow hcp sites. Upon adsorption, the net spin magnetic moment of the chemisorbed system decreases in each case compared to the bare surface. Difference charge density distributions clearly show that bonds between the surface Am atoms and the adatoms at each site is largely ionic in character. A study of the local density of states for O showed Am ($6d$)-Am ($5f$)-adatom($2p$) hybridizations at the top site electrons upon chemisorption, while at the bridge and hollow hcp sites the interactions are dominated by Am($6d$)-adatom($2p$). In the general, the $5f$ electron localization behavior of the Am atoms is primarily retained after chemisorption.

**Acknowledgments**

This work is supported by the Chemical Sciences, Geosciences and Biosciences Division, Office of Basic Energy Sciences, Office of Science, U. S. Department of Energy (Grant No. DE-FG02-03ER15409) and the Welch Foundation, Houston, Texas (Grant No. Y-1525). In addition to the supercomputing facilities at the University of Texas at Arlington, this research also used resources of the National Energy Research Scientific Computing Center, which is supported by the Office of Science of the U.S. Department of Energy under Contract No. DE-AC02-05CH11231. The authors also acknowledge the Texas Advanced Computing Center (TACC) at the University of Texas at Austin (http://www.tacc.utexas.edu) for providing computational resources.

**References**

1. J. J. Katz, G. T. Seaborg, and L. R. Morss, *The Chemistry of the Actinide Elements* (Chapman and Hall, 1986); L. R. Morss and J. Fuger, Eds. *Transuranium Elements: A Half Century* (American Chemical Society, Washington, D. C. 1992); L. R. Morss, N. M. Edelstein, and J. Fuger, Eds; J. J. Katz, Hon. Ed. *Chemistry of the Actinide and Transactinide Elements* (Springer, New York, 2006).

2. *Challenges in Plutonium Science,* Vol. I and II, Los Alamos Science, **26** (2000).

3. R. Haire, S. Heathman, M. Idiri, T. Le Bihan, and A. Lindbaum, Nuclear Materials Technology/Los Alamos National Laboratory, 3$^{rd}$/4$^{th}$ quarter 2003, p. 23.

4. *Fifty Years with Transuranium Elements*, Proceedings of the Robert A. Welch Foundation, October 22-23, 1990, Houston, Texas.

5. J. L. Sarrao, A. J. Schwartz, M. R. Antonio, P. C. Burns, R. G. Haire, and H. Nitsche, Eds. *Actinides 2005-Basic Science, Applications, and Technology,* Proceedings of the Materials Research Society, **893** (2005); K. J. M. Blobaum, E.A. Chandler, L. Havela, M. B. Maple, M. P. Neu, Eds. *Actinides 2006-Basic Science, Applications, and Technology,* Proceedings of the Materials Research Society, **986** (2006).

6. *The Elements beyond Uranium,* Glenn T. Seaborg and Walter D. Loveland, p. 17, (John Wiley & Sons, Inc. 1990).

7. S. Heathman, R. G. Haire, T. Le Bihan, A. Lindbaum, K. Litfin, Y. Méresse, and H. Libotte, Phys. Rev. Lett. **85**, 2961 (2000).

8. G. H. Lander and J. Fuger, Endeavour, **13**, 8, (1989).

9. A. J. Freeman and D. D. Koelling, in The *Actinides: Electronic Structure and Related Properties,* edited by A. J. Freeman and J. B. Darby, Jr. (Academic, New York, 1974).

10. B. Johansson, Phys. Rev. B. **11**, 2740 (1975).

11. H. L. Skriver, O. K. Andersen, and B. Johansson, Phys. Rev. Lett. **41**, 42 (1978).

12. J. R. Naegele, L. Manes, J. C. Spirlet, and W. Müller, Phys. Rev. Lett. **52**, 1834 (1984).

13. A. Lindbaum, S. Heathman, K. Litfin and Y. Méresse, Phys. Rev. B. **63**, 214101 (2001).

14. M. Pénicaud, J. Phys. Cond. Matt. **14,** 3575 (2002); *ibid*, **17**, 257 (2005).

15. S. Y. Savrasov, K. Haule, and G. Kotliar, Phys. Rev. Lett. **96**, 036404 (2006).

16. P. Sõderlind, R. Ahuja, O. Eriksson, B. Johansson, and J. M. Wills, Phys. Rev. B. **61**, 8119 (2000); P. Sõderlind and A. Landa, *ibid*, **72**, 024109 (2005).

17. P. G. Huray, S. E. Nave, and R. G. Haire, J. Less-Com. Met. **93**, 293 (1983).

18. T. Gouder, P. M. Oppeneer, F. Huber, F. Wastin, and J. Rebizant, Phys. Rev. B 72**,** 115122 (2005); L. E. Cox, J. W. Ward, and R. G. Haire, Phys. Rev. B **45**, 13239 (1992).

19. O. Eriksson and J. M. Wills, Phys. Rev. B **45**, 3198 (1992).

20. A. L. Kutepov, and S. G. Kutepova, J. Magn. Magn. Mat. **272-276,** e329 (2004).

21. A. Shick, L. Havela, J. Kolorenc, V. Drchal, T. Gouder, and P. M. Oppeneer, Phys. Rev. B **73**, 104415 (2006).

22. S. Y. Savrasov, G. Kotliar, and E. Abrahams, Nature **410,** 793 (2001); G. Kotliar and D. Vollhardt, Phys. Today **57**, 53 (2004); X. Dai, S. Y. Savrasov, G. Kotliar, A. Migliori, H. Ledbetter, and E. Abrahams, Science **300**, 953 (2003).

23. B. Johansson and A. Rosengren, Phys. Rev. B **11**, 2836 (1975).

24. J. L. Smith and R. G. Haire, Science **200**, 535 (1978).

25. J. C. Griveau, J. Rebizant, G. H. Lander, and G. Kotliar, Phys. Rev. Lett. **94,** 097002 (2005).

26. D. Gao and A. K. Ray, Eur. Phys. J. B, **50**, 497 (2006); MRS Fall 2005 Symp. Proc. **893**, 39 (2006); Surf. Sci., **600**, 4941 (2006); D. Gao and A. K. Ray, Eur. Phys. J. B **55**, 13 (2007) and references therein.

27. R. Atta-Fynn and A. K. Ray, Physica B, **392,** 112 (2007); Phys. Rev. B **75**, 195112 (2007) and references therein.

28. P. Hohenberg and W. Kohn, Phys. Rev. **136**, B864 (1964); W. Kohn and L. J. Sham, Phys. Rev. **140**, A1133 (1965).

29. J. P. Perdew, K. Burke, and M. Ernzerhof, Phys. Rev. Lett. **77**, 3865 (1996).

30. P. Blaha, K. Schwarz, G. K. H. Madsen, D. Kvasnicka, and J. Luitz, *WIEN2k, An Augmented Plane Wave Plus Local Orbitals Program for Calculating Crystal properties* (Vienna University of Technology, Austria, 2001).

31. D. D. Koelling and B. N. Harmon, J. Phys. C **10**, 3107 (1977).

32. J. Kunes, P. Novak, R. Schmid, P. Blaha, and K. Schwarz, Phys. Rev. B **64** 153102 (2001).

33. F. D. Murnaghan, Proc. Natl. Acad. Sci. USA **30**, 244 (1944).

34. R. W. G. Wyckoff, *Crystal Structure*s Volume 1 (Wiley, New York, 1963).

35. S. Heathman, R. G. Haire, T. Le Bihan, A. Lindbaum, K. Litfin, Y. Méresse, and H. Libotte, Phy. Rev. Lett. **85**, 2961 (2000); A. Lindbaum, S. Heathman, K. Litfin, Y. Méresse, R. G. Haire, T. Le Bihan, and H. Libotte, Phy. Rev. B **63**, 214101 (2001).

36. F. Wagner, Th. Laloyaux, and M. Scheffler, Phys. Rev. B **57**, 2102 (1998); J. L. F. Da Silva, C. Stampfl, and M. Scheffler, Surf. Sci. **600**, 703 (2006).

37. A. Kokalj, J. Mol. Graphics Modeling **17**, 176 (1999); code available from http://www.xcrysden.org

Table 1: Chemisorption energies $E_c$, distance of the adatom from the surface layer $R$, the distance of the adatom from the nearest neighbor Am atom $D_{Am-adatom}$ at both the NSOC and SOC levels of theory. $\Delta E_C = E_C(SOC) - E_C(NSOC)$ is the difference between the chemisorption energies at each adsorption site.

| Adatom | Site | $E_C$ (eV) (NSO) | $E_C$ (eV) (SO) | $R$ (Å) | $D_{Am-adatom}$ (Å) | $\Delta E_C$ (eV) |
|---|---|---|---|---|---|---|
| Hydrogen | Top | 2.272 | 2.377 | 2.122 | 2.122 | 0.105 |
|  | Bridge | 2.965 | 3.014 | 1.429 | 2.277 | 0.049 |
|  | Hcp | 3.136 | 3.217 | 1.196 | 2.371 | 0.081 |
| Oxygen | Top | 6.388 | 6.599 | 1.911 | 1.911 | 0.211 |
|  | Bridge | 8.204 | 8.368 | 1.164 | 2.121 | 0.164 |
|  | Hcp | 8.109 | 8.347 | 0.878 | 2.228 | 0.238 |

Table 2: Change in work function $\Delta\Phi = \Phi^{adatom/Am} - \Phi^{Am}$ (in eV) for both the NSOC and SOC levels of theory. $\Phi^{Am}$ = 2.906 eV and 2.989 eV respectively at the NSOC and SOC theoretical level.

| Theory | Site | Hydrogen | Oxygen |
|---|---|---|---|
| NSOC | Top | 1.149 | 1.343 |
|  | Bridge | 0.321 | 0.499 |
|  | Hcp | 0.156 | 0.388 |
| SOC | Top | 1.138 | 1.339 |
|  | Bridge | 0.319 | 0.477 |
|  | Hcp | 0.151 | 0.314 |

Table 3: $\mu_1$, $\mu_2$, $\mu_3$ are respectively the site projected spin magnetic moment for each Am atom for the surface layer, subsurface layer, and central layer. $\mu_{int}$ is the total spin magnetic moment in the interstitial region and $\mu_{tot}$ is the net (site + interstitial) magnetic moment per atom. Spin moments are quoted for SOC calculations.

|  | Site | $\mu_1 (\mu_B)$ | $\mu_2 (\mu_B)$ | $\mu_3 (\mu_B)$ | $\mu_{int} (\mu_B)$ | $\mu_{tot} (\mu_B/\text{Am atom})$ |
|---|---|---|---|---|---|---|
| Bare Slab |  | 5.81, 5.81<br>5.81, 5.81 | -5.67, -5.67<br>-5.67, -5.67 | 5.69, 5.69<br>5.69, 5.69 | 9.61 | 2.34 |
| Hydrogen | Top | **5.67**, 5.80<br>5.80, 5.80 | -5.67, -5.67<br>-5.67, -5.67 | 5.68, 5.68<br>5.68, 5.68 | 8.95 | 2.30 |
| Hydrogen | Bridge | **5.73**, **5.73**<br>5.80, 5.80 | -5.67, -5.67<br>-5.67, -5.67 | 5.69, 5.69<br>5.69, 5.69 | 8.69 | 2.28 |
| Hydrogen | Hcp | **5.75**, **5.75**<br>**5.75**, 5.79 | -5.67, -5.67<br>-5.67, -5.67 | 5.69, 5.69<br>5.69, 5.69 | 8.70 | 2.28 |
| Oxygen | Top | **5.19**, 5.81<br>5.81, 5.81 | -5.66, -5.66<br>-5.66, -5.66 | 5.68, 5.68<br>5.68, 5.68 | 8.29 | 2.23 |
| Oxygen | Bridge | **5.54**, **5.54**<br>5.81, 5.81 | -5.68, -5.68<br>-5.67, -5.67 | 5.69, 5.69<br>5.69, 5.69 | 8.14 | 2.22 |
| Oxygen | Hcp | **5.65**, **5.65**<br>**5.65**, 5.81 | -5.68, -5.67<br>-5.67, -5.67 | 5.68, 5.68<br>5.68, 5.68 | 8.48 | 2.25 |

Table 4: Partial charges inside muffin tin spheres before adsorption ($Q_B$), after adsorption ($Q_A$), and difference in partial charges $\Delta Q = Q_A - Q_B$ at the **top** site for hydrogen at the SOC level of theory. The surface layer atoms with which the adatom interacts are given in bold fonts.

| Atom/Layer | Partial charges in muffin-tin | | | | | | $\Delta Q = Q_A - Q_B$ | | |
| --- | --- | --- | --- | --- | --- | --- | --- | --- | --- |
| | Before adsorption $Q_B$ | | | After adsorption $Q_A$ | | | | | |
| | H $s$ | Am $d$ | Am $f$ | H $s$ | Am $d$ | Am $f$ | H $s$ | Am $d$ | Am $f$ |
| Hydrogen (top) | 0.41 | | | 0.57 | | | 0.16 | | |
| Am surface layer | | 0.27 | 5.85 | | 0.26 | 5.87 | | -0.01 | 0.02 |
| | | 0.27 | 5.85 | | 0.26 | 5.87 | | -0.01 | 0.02 |
| | | 0.27 | 5.85 | | **0.31** | **5.76** | | 0.04 | -0.09 |
| | | 0.27 | 5.85 | | 0.26 | 5.87 | | -0.01 | 0.02 |
| | | | | | | | | | |
| Am subsurface layer | | 0.32 | 5.75 | | 0.32 | 5.75 | | 0.00 | 0.00 |
| | | 0.32 | 5.75 | | 0.32 | 5.76 | | 0.00 | 0.01 |
| | | 0.32 | 5.75 | | 0.32 | 5.76 | | 0.00 | 0.01 |
| | | 0.32 | 5.75 | | 0.32 | 5.76 | | 0.00 | 0.01 |
| | | | | | | | | | |
| Am central layer | | 0.32 | 5.77 | | 0.32 | 5.76 | | -0.01 | -0.01 |
| | | 0.32 | 5.77 | | 0.32 | 5.76 | | -0.01 | -0.01 |
| | | 0.32 | 5.77 | | 0.32 | 5.76 | | -0.01 | -0.01 |
| | | 0.32 | 5.77 | | 0.32 | 5.76 | | -0.01 | -0.01 |

Table 5: Partial charges inside muffin tin spheres before adsorption ($Q_B$), after adsorption ($Q_A$), and difference in partial charges $\Delta Q = Q_A - Q_B$ at the **bridge** site for hydrogen at the SOC level of theory. The surface layer atoms with which the adatom interacts are given in bold fonts.

| Atom/Layer | Partial charges in muffin-tin | | | | | | $\Delta Q = Q_A - Q_B$ | | |
| --- | --- | --- | --- | --- | --- | --- | --- | --- | --- |
| | Before adsorption $Q_B$ | | | After adsorption $Q_A$ | | | | | |
| | H $s$ | Am $d$ | Am $f$ | H $s$ | Am $d$ | Am $f$ | H $s$ | Am $d$ | Am $f$ |
| Hydrogen (bridge) | 0.41 | | | 0.59 | | | 0.18 | | |
| Am surface layer | | 0.27 | 5.85 | | **0.28** | **5.82** | | 0.01 | -0.03 |
| | | 0.27 | 5.85 | | **0.28** | **5.82** | | 0.01 | -0.03 |
| | | 0.27 | 5.85 | | 0.26 | 5.86 | | -0.01 | 0.01 |
| | | 0.27 | 5.85 | | 0.26 | 5.86 | | -0.01 | 0.01 |
| | | | | | | | | | |
| Am subsurface layer | | 0.32 | 5.75 | | 0.32 | 5.76 | | 0.00 | 0.01 |
| | | 0.32 | 5.75 | | 0.32 | 5.75 | | 0.00 | 0.00 |
| | | 0.32 | 5.75 | | 0.32 | 5.75 | | 0.00 | 0.00 |
| | | 0.32 | 5.75 | | 0.32 | 5.75 | | 0.00 | 0.00 |
| | | | | | | | | | |
| Am central layer | | 0.32 | 5.77 | | 0.32 | 5.77 | | 0.00 | 0.00 |
| | | 0.32 | 5.77 | | 0.32 | 5.77 | | 0.00 | 0.00 |
| | | 0.32 | 5.77 | | 0.32 | 5.77 | | 0.00 | 0.00 |
| | | 0.32 | 5.77 | | 0.32 | 5.77 | | 0.00 | 0.00 |

Table 6: Partial charges inside muffin tin spheres before adsorption ($Q_B$), after adsorption ($Q_A$), and difference in partial charges $\Delta Q = Q_A - Q_B$ at the **hcp** site for hydrogen at the SOC level of theory. The surface layer atoms with which the adatom interacts are given in bold fonts.

| Atom/Layer | Partial charges in muffin-tin | | | | | | $\Delta Q = Q_A - Q_B$ | | |
|---|---|---|---|---|---|---|---|---|---|
| | Before adsorption $Q_B$ | | | After adsorption $Q_A$ | | | | | |
| | H $s$ | Am $d$ | Am $f$ | H $s$ | Am $d$ | Am $f$ | H $s$ | Am $d$ | Am $f$ |
| Hydrogen (hcp) | 0.41 | | | 0.59 | | | 0.18 | | |
| Am surface layer | | 0.27 | 5.85 | | **0.28** | **5.83** | | 0.01 | -0.02 |
| | | 0.27 | 5.85 | | **0.28** | **5.83** | | 0.01 | -0.02 |
| | | 0.27 | 5.85 | | 0.27 | 5.85 | | 0.00 | 0.00 |
| | | 0.27 | 5.85 | | **0.28** | **5.83** | | 0.01 | -0.02 |
| | | | | | | | | | |
| Am subsurface layer | | 0.32 | 5.75 | | 0.32 | 5.75 | | 0.00 | 0.00 |
| | | 0.32 | 5.75 | | 0.32 | 5.75 | | 0.00 | 0.00 |
| | | 0.32 | 5.75 | | 0.32 | 5.75 | | 0.00 | 0.00 |
| | | 0.32 | 5.75 | | 0.32 | 5.75 | | 0.00 | 0.00 |
| | | | | | | | | | |
| Am central layer | | 0.32 | 5.77 | | 0.32 | 5.77 | | 0.00 | 0.00 |
| | | 0.32 | 5.77 | | 0.32 | 5.77 | | 0.00 | 0.00 |
| | | 0.32 | 5.77 | | 0.32 | 5.77 | | 0.00 | 0.00 |
| | | 0.32 | 5.77 | | 0.32 | 5.77 | | 0.00 | 0.00 |

Table 7: Partial charges inside muffin tin spheres before adsorption ($Q_B$), after adsorption ($Q_A$), and difference in partial charges $\Delta Q = Q_A - Q_B$ at the **top** site for oxygen at the SOC level of theory. The surface layer atoms with which the adatom interacts are given in bold fonts.

| Atom/Layer | Partial charges in muffin-tin | | | | | | $\Delta Q = Q_A - Q_B$ | | |
| --- | --- | --- | --- | --- | --- | --- | --- | --- | --- |
| | Before adsorption $Q_B$ | | | After adsorption $Q_A$ | | | | | |
| | O $p$ | Am $d$ | Am $f$ | O $p$ | Am $d$ | Am $f$ | O $p$ | Am $d$ | Am $f$ |
| Oxygen (top) | 2.17 | | | 2.38 | | | 0.21 | | |
| Am surface layer | | 0.27 | 5.85 | | 0.25 | 5.88 | | -0.02 | 0.03 |
| | | 0.27 | 5.85 | | 0.25 | 5.88 | | -0.02 | 0.03 |
| | | 0.27 | 5.85 | | **0.48** | **5.59** | | 0.21 | -0.26 |
| | | 0.27 | 5.85 | | 0.25 | 5.88 | | -0.02 | 0.03 |
| | | | | | | | | | |
| Am subsurface layer | | 0.32 | 5.75 | | 0.32 | 5.75 | | 0.00 | 0.00 |
| | | 0.32 | 5.75 | | 0.32 | 5.76 | | 0.00 | 0.01 |
| | | 0.32 | 5.75 | | 0.32 | 5.76 | | 0.00 | 0.01 |
| | | 0.32 | 5.75 | | 0.32 | 5.76 | | 0.00 | 0.01 |
| | | | | | | | | | |
| Am central layer | | 0.32 | 5.77 | | 0.32 | 5.76 | | 0.00 | -0.01 |
| | | 0.32 | 5.77 | | 0.32 | 5.76 | | 0.00 | -0.01 |
| | | 0.32 | 5.77 | | 0.32 | 5.76 | | 0.00 | -0.01 |
| | | 0.32 | 5.77 | | 0.32 | 5.76 | | 0.00 | -0.01 |
| | | | | | | | | | |

Table 8: Partial charges inside muffin tin spheres before adsorption ($Q_B$), after adsorption ($Q_A$), and difference in partial charges $\Delta Q = Q_A - Q_B$ at the **bridge** site for oxygen at the SOC level of theory. The surface layer atoms with which the adatom interacts are given in bold fonts.

| Atom/Layer | Partial charges in muffin-tin | | | | | | $\Delta Q = Q_A - Q_B$ | | |
|---|---|---|---|---|---|---|---|---|---|
| | Before adsorption $Q_B$ | | | After adsorption $Q_A$ | | | | | |
| | O $p$ | Am $d$ | Am $f$ | O $p$ | Am $d$ | Am $f$ | O $p$ | Am $d$ | Am $f$ |
| Oxygen (bridge) | 2.17 | | | 2.42 | | | 0.25 | | |
| Am surface layer | | 0.27 | 5.85 | | **0.33** | **5.73** | | 0.06 | -0.12 |
| | | 0.27 | 5.85 | | **0.33** | **5.74** | | 0.06 | -0.11 |
| | | 0.27 | 5.85 | | 0.25 | 5.88 | | -0.02 | 0.03 |
| | | 0.27 | 5.85 | | 0.25 | 5.88 | | -0.02 | 0.03 |
| Am subsurface layer | | 0.32 | 5.75 | | 0.32 | 5.76 | | 0.00 | 0.01 |
| | | 0.32 | 5.75 | | 0.32 | 5.75 | | 0.00 | 0.00 |
| | | 0.32 | 5.75 | | 0.32 | 5.75 | | 0.00 | 0.00 |
| | | 0.32 | 5.75 | | 0.32 | 5.75 | | 0.00 | 0.00 |
| Am central layer | | 0.32 | 5.77 | | 0.32 | 5.76 | | 0.00 | 0.01 |
| | | 0.32 | 5.77 | | 0.32 | 5.76 | | 0.00 | 0.01 |
| | | 0.32 | 5.77 | | 0.32 | 5.76 | | 0.00 | 0.01 |
| | | 0.32 | 5.77 | | 0.32 | 5.76 | | 0.00 | 0.01 |

Table 9: Partial charges inside muffin tin spheres before adsorption ($Q_B$), after adsorption ($Q_A$), and difference in partial charges $\Delta Q = Q_A - Q_B$ at the **hcp** site for oxygen at the SOC level of theory. The surface layer atoms with which the adatom interacts are given in bold fonts.

| Atom/Layer | Partial charges in muffin-tin | | | | | | ΔQ = Q$_A$ − Q$_B$ | | |
|---|---|---|---|---|---|---|---|---|---|
| | Before adsorption Q$_B$ | | | After adsorption Q$_A$ | | | | | |
| | O p | Am d | Am f | O p | Am d | Am f | O p | Am d | Am f |
| Oxygen (hcp) | 2.17 | | | 2.44 | | | 0.27 | | |
| Am surface layer | | 0.27 | 5.85 | | **0.29** | **5.79** | | 0.02 | -0.06 |
| | | 0.27 | 5.85 | | **0.29** | **5.79** | | 0.02 | -0.06 |
| | | 0.27 | 5.85 | | 0.27 | 5.86 | | 0.00 | 0.01 |
| | | 0.27 | 5.85 | | **0.29** | **5.79** | | 0.02 | -0.06 |
| | | | | | | | | | |
| Am subsurface layer | | 0.32 | 5.75 | | 0.32 | 5.76 | | 0.00 | 0.01 |
| | | 0.32 | 5.75 | | 0.32 | 5.75 | | 0.00 | 0.00 |
| | | 0.32 | 5.75 | | 0.32 | 5.75 | | 0.00 | 0.00 |
| | | 0.32 | 5.75 | | 0.32 | 5.75 | | 0.00 | 0.00 |
| | | | | | | | | | |
| Am central layer | | 0.32 | 5.77 | | 0.32 | 5.76 | | 0.00 | -0.01 |
| | | 0.32 | 5.77 | | 0.32 | 5.76 | | 0.00 | -0.01 |
| | | 0.32 | 5.77 | | 0.32 | 5.76 | | 0.00 | -0.01 |
| | | 0.32 | 5.77 | | 0.32 | 5.76 | | 0.00 | -0.01 |

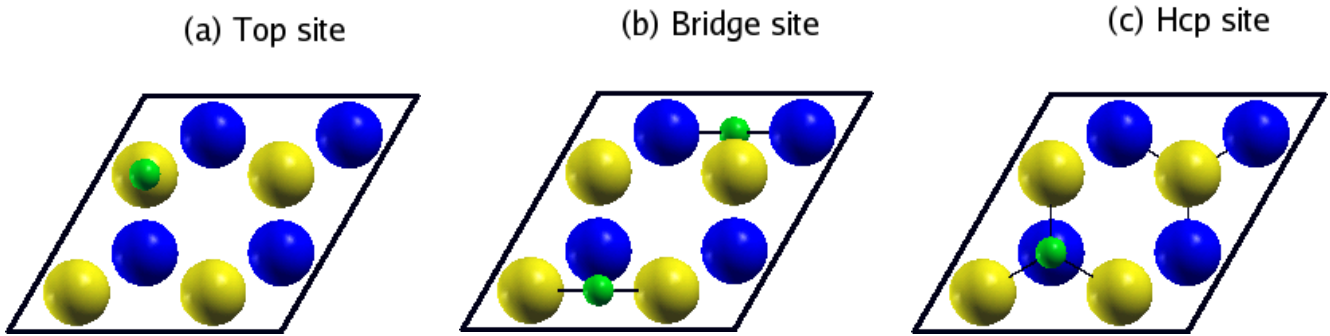

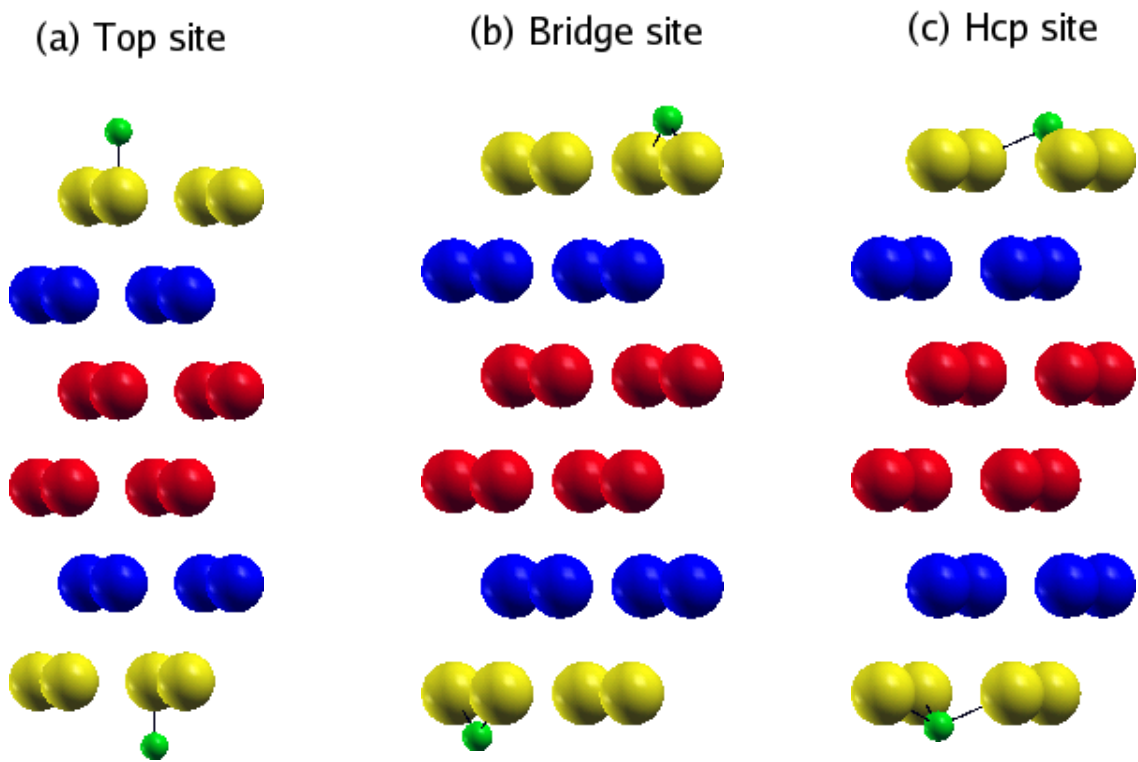

FIG. 1 (Color online) Top and side view illustrations of the three high-symmetry adsorption sites for the six-layer dhcp-Am(0001) slab with a 0.25 ML adlayer coverage: (a) one-fold top site; (b) two-fold bridge site; (c) three-fold hcp site. Atoms are colored to distinguish between the layers. Surface, subsurface, and central layers are colored gold, blue, and red, respectively. Adatom is colored green.

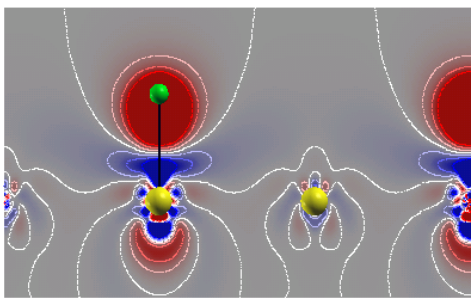
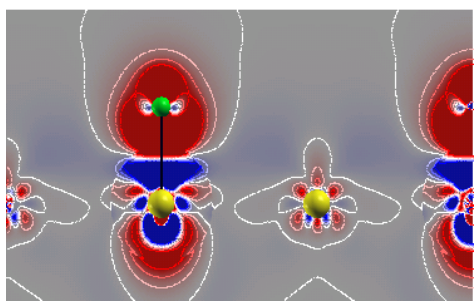
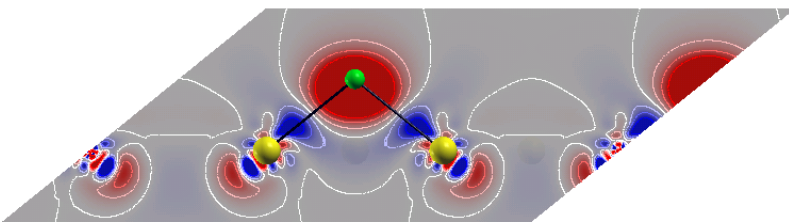
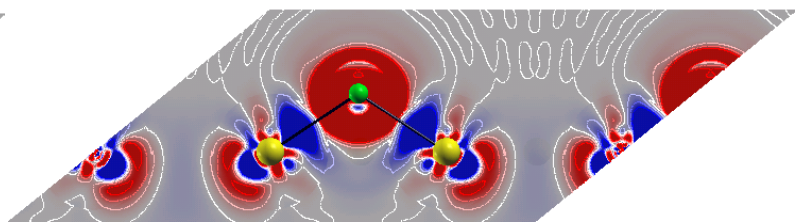
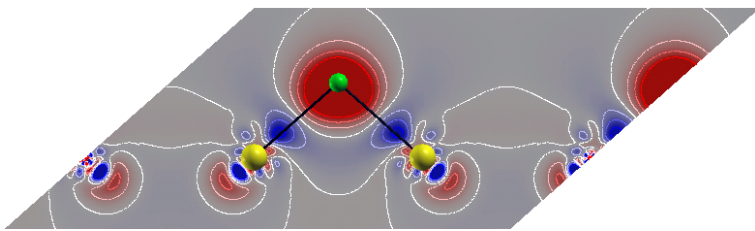
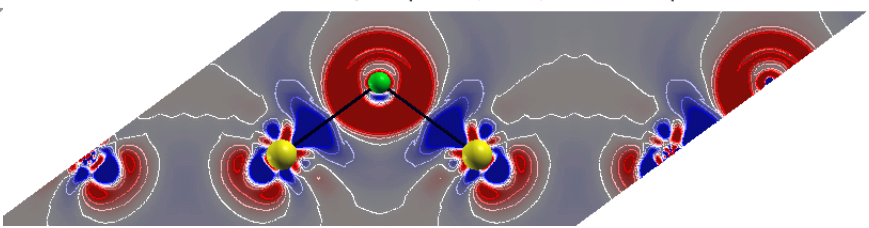
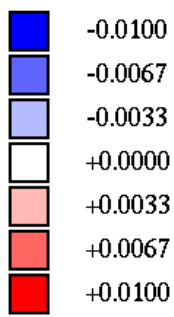

FIG. 2.(Color online) Difference charge density distributions Δn(r) for O and H chemisorbed on the dhcp-Am(0001) surface. Charge densities were computed in a plane passing through the adatom and two neighboring Am atoms. The scale used is shown at the bottom. Red (positive) denotes regions of charge accumulation and blue (negative) denotes regions of charge loss. Adatoms are colored green and Am atoms are colored gold.

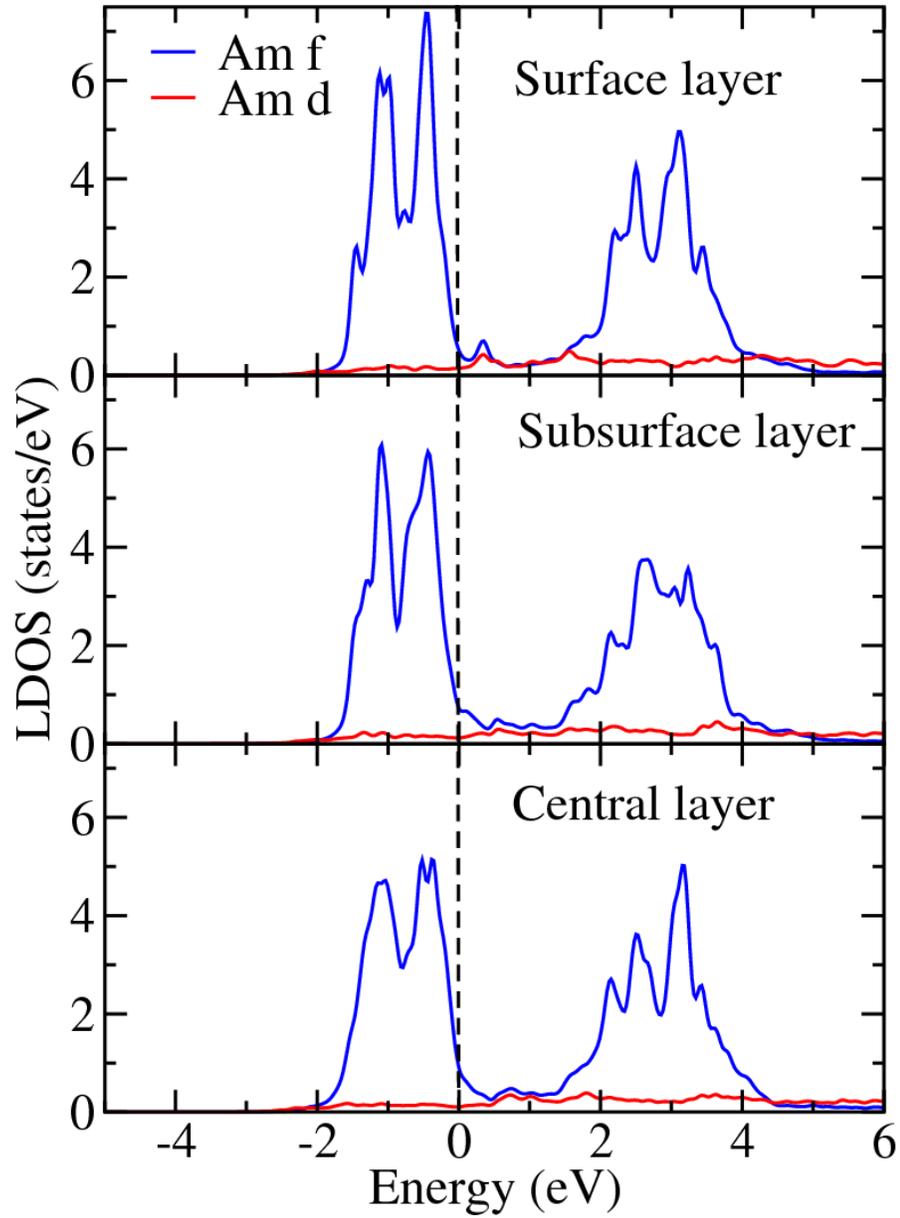

FIG. 3 (Color Online) $d$ and $f$ LDOS curves inside the muffin-tins for each layer of the bare dhcp-Am(0001) slab. Vertical line through E=0 is the Fermi level. LDOS correspond to calculations with SOC.

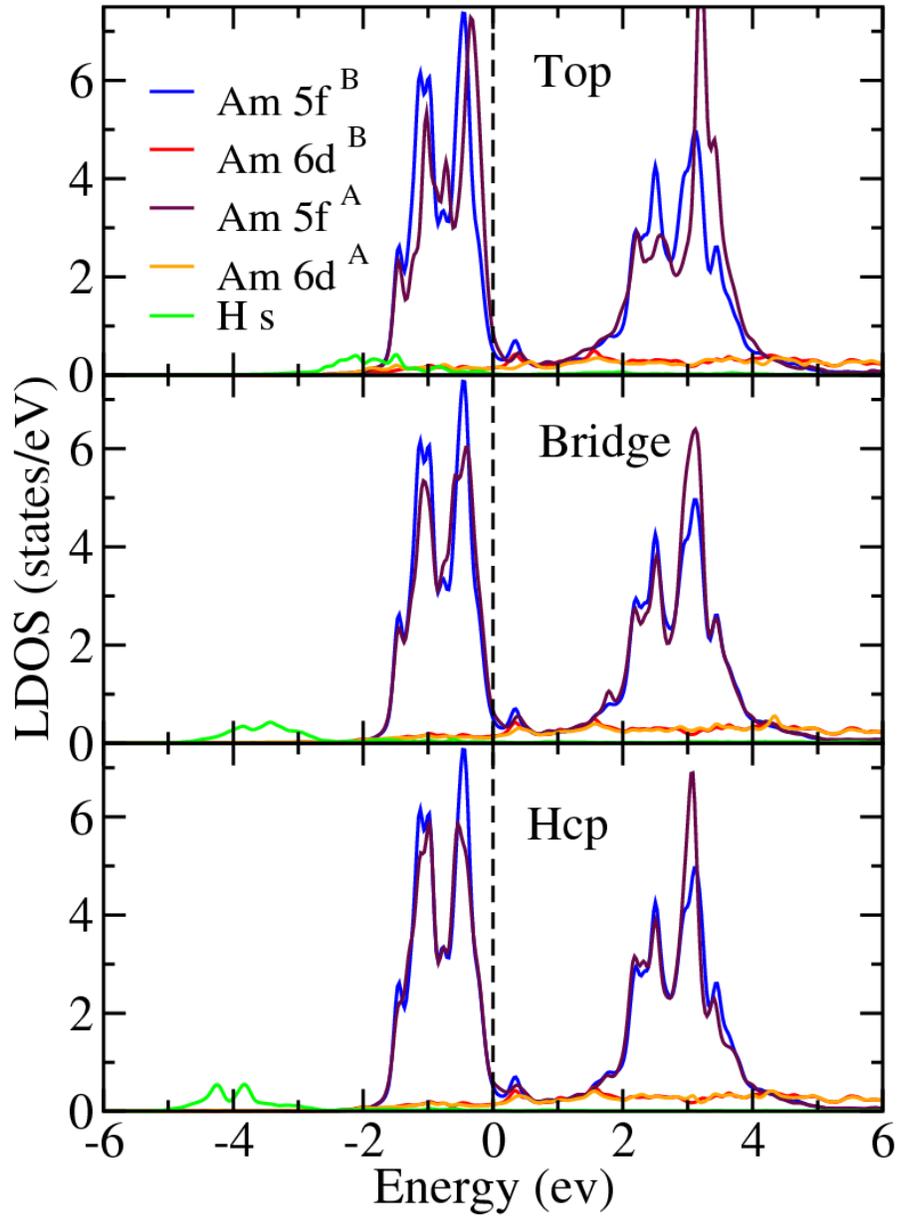

FIG. 4 (Color Online) $d$ and $f$ LDOS curves inside the muffin-tins for the Am atoms on the surface layer and $s$ LDOS curves for H adatom. Vertical line through E=0 is the Fermi level. LDOS correspond to calculations with SOC. Superscripts $B$ and $A$ refer to Am $d$ and $f$ surface layer LDOS before (top panel in FIG. 3) and after adsorption, respectively.

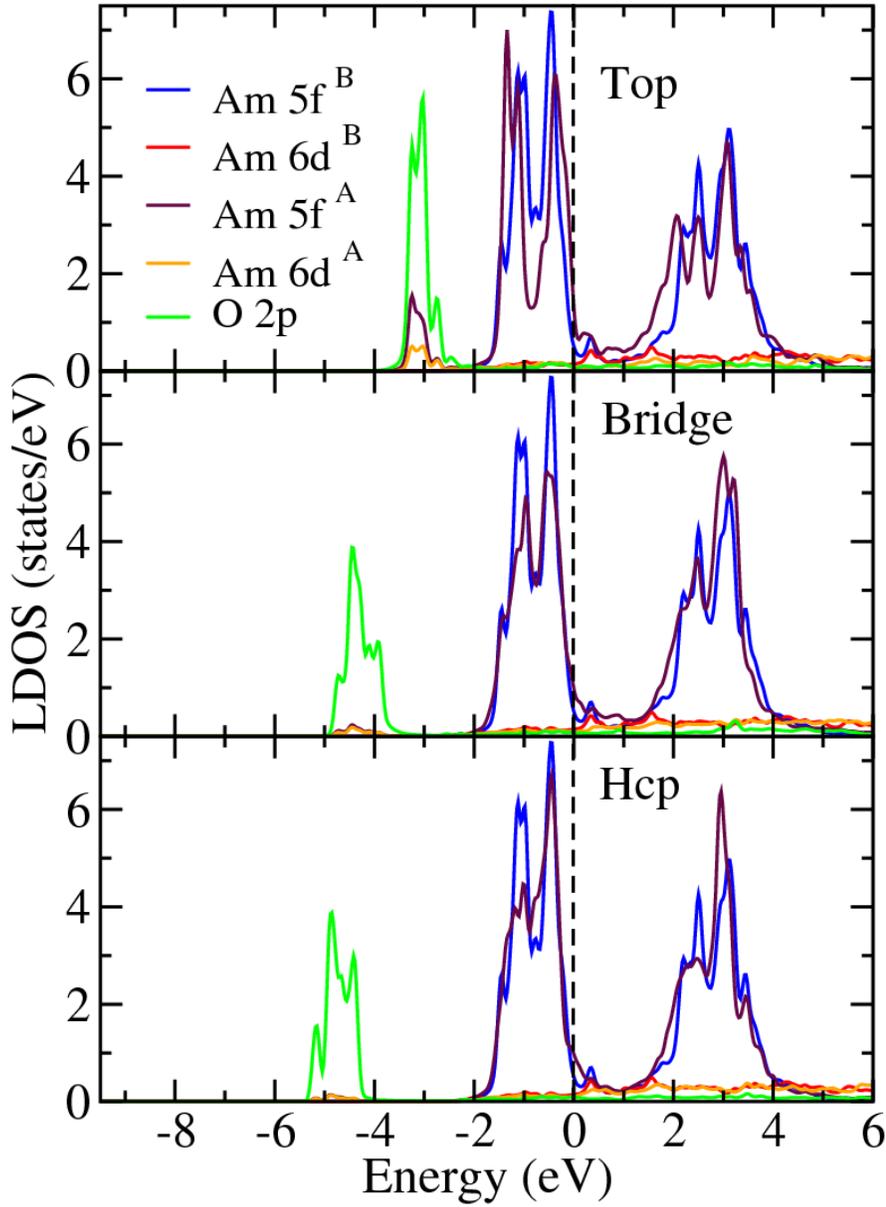

FIG. 5 (Color Online) $d$ and $f$ LDOS curves inside the muffin-tins for the Am atoms on the surface layer and $p$ LDOS curves for O adatom. Vertical line through E=0 is the Fermi level. LDOS correspond to calculations with SOC. Superscripts $B$ and $A$ refer to Am $d$ and $f$ surface layer LDOS before (top panel in FIG. 3) and after adsorption, respectively.